# Magnetisation Studies of Geometrically Frustrated Antiferromagnets $SrLn_2O_4$, with $Ln$ = Er, Dy and Ho

Thomas J. Hayes[1], Olga Young[1]*, Geetha Balakrishnan[1] and Oleg A. Petrenko[1]

[1] *Department of Physics, University of Warwick, Coventry CV4 7AL, United Kingdom*

We present the results of susceptibility $\chi(T)$ and magnetisation $M(H)$ measurements performed on single crystal samples of the rare-earth oxides $SrLn_2O_4$ ($Ln$ = Er, Dy and Ho). The measurements reveal the presence of magnetic ordering transition in $SrHo_2O_4$ at 0.62 K and confirm that $SrEr_2O_4$ orders magnetically at 0.73 K, while in $SrDy_2O_4$ such a transition is absent down to at least 0.5 K. The observed ordering temperatures are significantly lower than the Curie-Weiss temperatures, $\theta_{CW}$, obtained from the high-temperature linear fits to the $1/\chi(T)$ curves, which implies that these materials are subject to geometric frustration. Strong anisotropy found in the $\chi(T)$ curves for a field applied along the different crystallographic directions is also evident in the $M(H)$ curves measured both above and below the ordering temperatures. For all three compounds the magnetisation plateaux at approximately one third of the magnetisation saturation values can be seen for certain directions of applied field, which is indicative of field-induced stabilisation of a collinear *two-up one-down* structure.

KEYWORDS: geometric frustration, antiferromagnet, magnetic anisotropy, magnetisation plateau

## 1. Introduction

The term *geometric frustration* is usually used to describe the inability of a magnetic system to satisfy all the competing interactions and to form a unique ground state as a result of structural constraints.[1,2] Frustration frequently leads to reduced critical temperatures and highly degenerate ground states due to residual entropy at low temperatures. In the presence of antiferromagnetic exchange interactions, magnets based on triangles give rise to geometric frustration.[3] The Kagome,[4,5] pyrochlore[6,7] and garnet[8,9] lattices are all well-known and much studied structures where frustration plays a fundamental role in establishing the low temperature properties.

With the search for new frustrated magnets intensifying, the number of systems known to display the effects of geometric frustration is constantly growing. A particularly interesting structure in this respect is the so-called honeycomb lattice, where in an ideal two-dimensional case each magnetic atom is linked to three equally-spaced neighbours 120° apart. The honeycomb lattice is bipartite as it is made up of a plane network of edge sharing hexagons. It would only be frustrated if further neighbour interactions are considered, but it has received substantial theoretical interest since it has the smallest possible coordination number in two-dimensions.[10,11] This implies that quantum fluctuations are expected to have a much larger effect on the stability of Néel order than, for example, for a square or triangular lattice. Recently, a number of such structures have been identified experimentally, including the spin-$\frac{1}{2}$ compound $InCu_{2/3}V_{1/3}O_3$,[12,13] the spin-$\frac{3}{2}$ systems $Bi_3Mn_4O_{12}(NO_3)$[14,15] and $\beta$-$CaCr_2O_4$,[16,17] as well as other systems with larger spins like $BaLn_2O_4$,[18] $EuLn_2O_4$[19] and $SrLn_2O_4$[20] ($Ln$ = rare-earth ion).

*Corresponding author. E-mail address: o.young@warwick.ac.uk

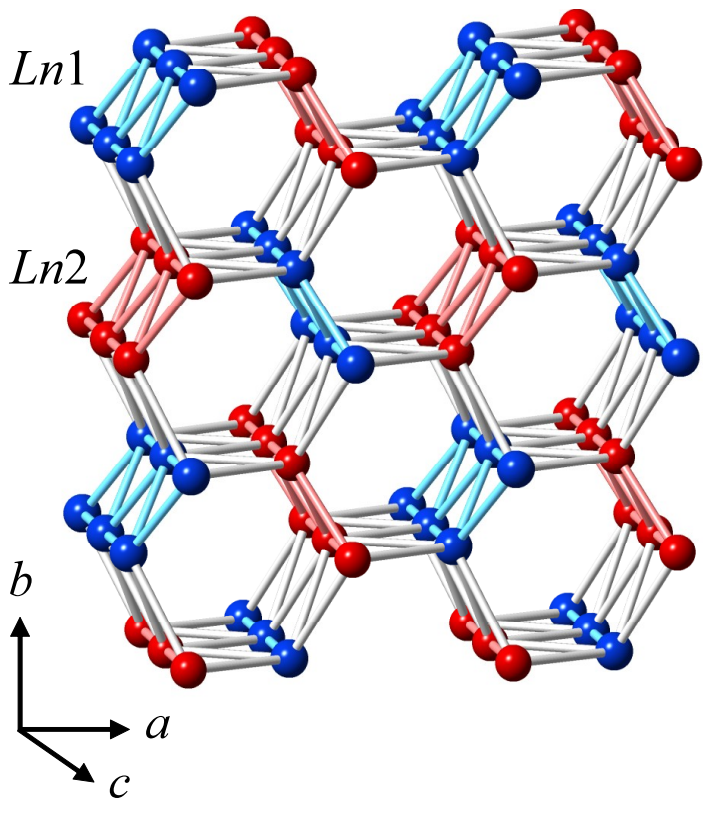

**Fig. 1.** (Colour online) Magnetic sublattice of $SrLn_2O_4$ compounds, with the two crystallographically inequivalent positions of the rare-earth $Ln$ ions shown in different colours. When viewed along the $c$ axis, honeycombs of the $Ln^{3+}$ ions are visible. Zigzag ladders running along the $c$ axis connect the honeycomb layers and give rise to geometric frustration.

In this paper we report on an investigation of the temperature dependence of the magnetic susceptibility as well as the field dependence of the magnetisation in three compounds from the $SrLn_2O_4$ family, with $Ln$ = Er, Dy and Ho.

The $SrLn_2O_4$ compounds crystallise in the form of calcium ferrite,[21] with the space group $Pnam$. The magnetic $Ln$ ions are linked in a network of triangles and hexagons, as shown in Fig. 1, with two crystallographically inequivalent sites for the rare-earth ions shown in red and blue. Since the distances between the magnetic ions which form the hexagons are not equal, varying, for example in $SrEr_2O_4$ from 3.478 Å to 4.019 Å,[22] the honeycomb structure is not perfect and should be described as "modified" or "distorted." It should also be noted that the zigzag triangular ladders formed by the $Ln$ ions are in fact magnetically equivalent to one-dimencional chains with first- and second-nearest-neighbour interactions.

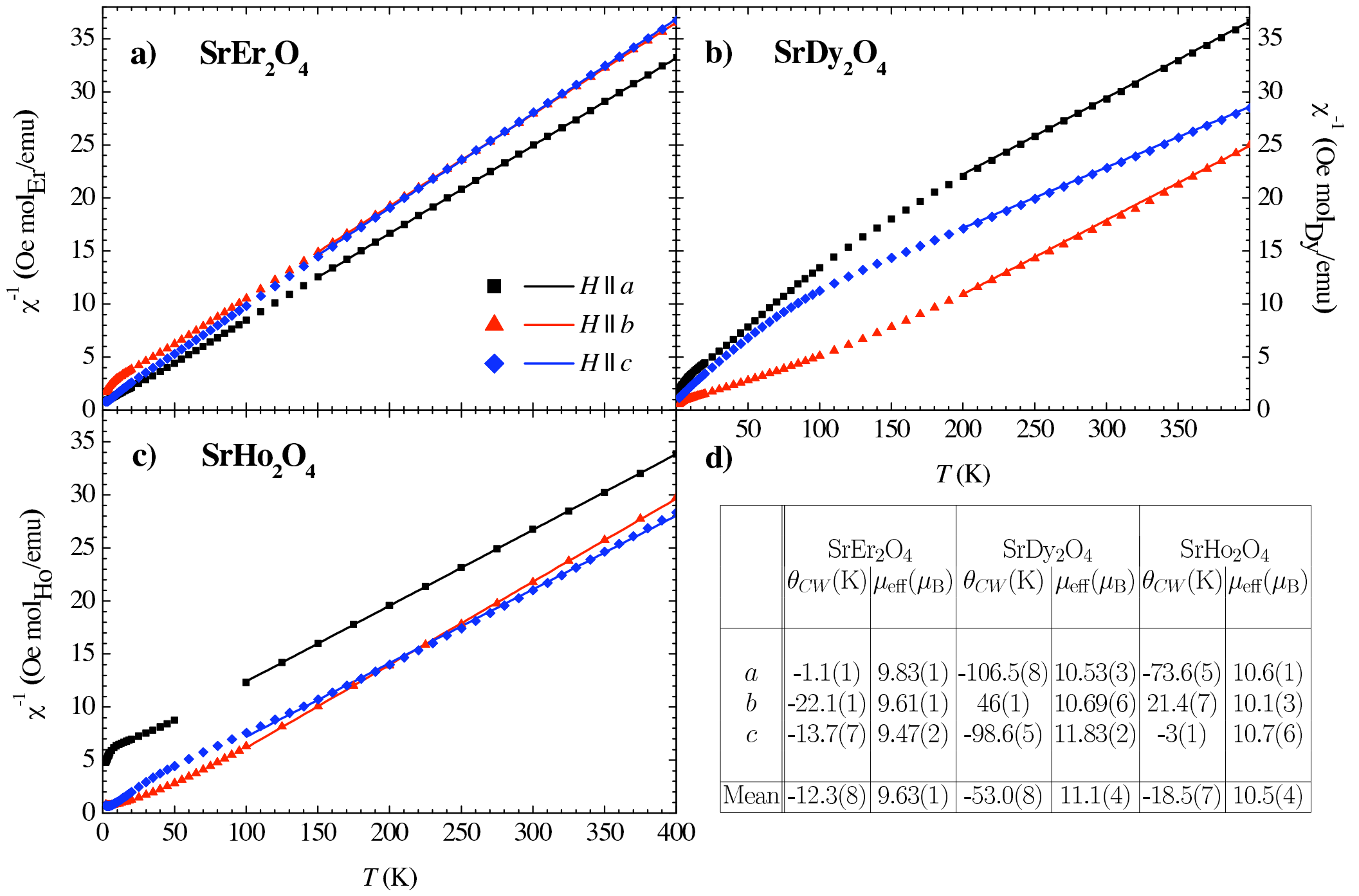

| | $SrEr_2O_4$ $\theta_{CW}$(K) | $SrEr_2O_4$ $\mu_{\rm eff}(\mu_B)$ | $SrDy_2O_4$ $\theta_{CW}$(K) | $SrDy_2O_4$ $\mu_{\rm eff}(\mu_B)$ | $SrHo_2O_4$ $\theta_{CW}$(K) | $SrHo_2O_4$ $\mu_{\rm eff}(\mu_B)$ |
|---|---|---|---|---|---|---|
| *a* | -1.1(1) | 9.83(1) | -106.5(8) | 10.53(3) | -73.6(5) | 10.6(1) |
| *b* | -22.1(1) | 9.61(1) | 46(1) | 10.69(6) | 21.4(7) | 10.1(3) |
| *c* | -13.7(7) | 9.47(2) | -98.6(5) | 11.83(2) | -3(1) | 10.7(6) |
| Mean | -12.3(8) | 9.63(1) | -53.0(8) | 11.1(4) | -18.5(7) | 10.5(4) |

**Fig. 2.** (Colour online) Reciprocal of the molar susceptibility versus temperature curves obtained in an applied field of 1 kOe over the temperature range 2-400 K for three crystals: (a) $SrEr_2O_4$, (b) $SrDy_2O_4$, (c) $SrHo_2O_4$. Lines are least-squares regression fits to the data using the Curie-Weiss model. (d) A table summarising the values of the Curie-Weiss constant, $\theta_{CW}$, and the effective moment per magnetic ion, $\mu_{\rm eff}$, from the fits.

The presence of geometric frustration can often be seen in bulk property measurements as the appearance of phase transitions at temperatures much lower than expected from the strength of the exchange interactions. Previous magnetic susceptibility measurements on powder samples of $SrLn_2O_4$ have revealed a disparity between the measured Curie-Weiss constant $\theta_{CW}$ and the lack of long-range order (LRO) down to 1.8 K.[22] The extended temperature range of our measurements, down to 0.5 K, allowed for the observation of a magnetic ordering transition in $SrHo_2O_4$ at 0.62 K. The measurements have also confirmed that $SrEr_2O_4$ orders magnetically at 0.73 K ($T_N$ of 0.75 K was previously reported from powder neutron diffraction and heat capacity measurements[23]). In $SrDy_2O_4$, however, no transition is observed down to 0.5 K.

Magnetisation curves $M(H)$ have revealed field-induced phase transitions at low temperatures for all three compounds. For all three crystals the magnetisation plateaux at approximately one third of the saturation values can be seen for certain directions of applied field, which is usually indicative of the stabilisation of a collinear *two-spins-up one-spin-down* (*uud*) structure.

## 2. Experimental Details

Single crystal samples, used for all of the measurements, were grown by the floating zone method, with details of the procedure used reported elsewhere.[29] The samples were aligned along the principal crystal axes to within an estimated accuracy of 2° using the backscattering x-ray diffraction Laue technique. The samples were cut into thin rectangular plates that weighed around 10 mg. In order to minimize the effects of demagnetisation, the field was applied parallel to the longest side of the plates. Corrections to the data to take into account this effect were made following Aharoni,[30] and the demagnetising field was found always to be less than 5% (typically 2%) of the applied field.

The temperature dependence of the magnetic susceptibility $\chi(T)$ up to 400 K and the field dependence of magnetisation $M(H)$ up to 70 kOe were measured using a Quantum Design SQUID magnetometer (MPMS5S), with temperatures down to 0.5 K achieved using an iQuantum (IQ2000-AGHS-2RSO $^3$He system) refrigerator insert.[31]

## 3. Experimental Results and Discussion

### *3.1 Temperature dependence of the magnetic susceptibility*

#### *3.1.1 High-temperature limit*

The magnetic susceptibility versus temperature was measured for fields applied along each of the three principal axes for single crystal samples of $SrEr_2O_4$, $SrHo_2O_4$ and $SrDy_2O_4$. The high-temperature dependence of the inverse susceptibility is presented in Fig. 2 for all of the three compounds. In the high-temperature range, where the $1/\chi(T)$ curves follow a paramagnetic Curie-Weiss behaviour, the data are fitted with least-squares regression lines, in the temperature range 150 K - 400 K for $SrEr_2O_4$; 200 K - 400 K for $SrDy_2O_4$; and 100 K - 400 K for $SrHo_2O_4$. The parameters of these fits, the Curie-Weiss constants, and the effective moments per magnetic ion ($\theta_{CW}$ and $\mu_{\rm eff}$) are also shown, in the table in Fig. 2d. From the graphs it is immediately obvious that there is a large anisotropy in the Curie-Weiss temperatures (some of which even change sign) for the three principal crystal directions of these compounds. The average values, however, are in general agreement with those presented by Karunadasa *et al.* for polycrystalline Sr$Ln_2$O$_4$ samples.[22)]

The observed high-temperature susceptibly curves imply that the magnetic anisotropy in SrEr$_2$O$_4$ is of easy-axis type with the $a$ axis being the easy direction for magnetisation. The situation changes dramatically, however, at lower temperatures, where $\chi_{H\|c}$ becomes similar in value to $\chi_{H\|a}$, while $\chi_{H\|b}$ still remains the lowest. In SrHo$_2$O$_4$ the anisotropy is of easy-plane type for the entire temperature range studied with the $a$ axis being the hard direction for magnetisation. In SrDy$_2$O$_4$, at low-temperature the anisotropy is clearly of easy-axis type with the $b$ axis being the easy direction for magnetisation, while during warming the $\chi_{H\|a}$ and $\chi_{H\|c}$ become more and more different and at temperatures above approximately 100 K the anisotropy cannot be classified as either easy-axis or easy-plane type. The anomalous behaviour of the susceptibility in SrDy$_2$O$_4$ for $H \parallel c$ is evident from the estimated value of the effective moment of the Dy$^{3+}$ ion, $\mu_{\rm eff} = 11.83(2)\ \mu_{\rm B}$, which exceeds significantly the expected value of 10.6 $\mu_{\rm B}$.[32)]

On close inspection it is evident that $1/\chi(T)$ curves in SrDy$_2$O$_4$ behave only approximately linearly with temperature for all three directions of the applied field and that both $\theta_{CW}$ and $\mu_{\rm eff}$ are influenced significantly by the range of temperature selected for fitting. In order to avoid possible ambiguity, we have prepared a powder sample from single crystals of SrDy$_2$O$_4$ and remeasured the temperature dependence of the susceptibility. The value of the effective moment of the Dy$^{3+}$ ion $\mu_{\rm eff} = 10.5\ \mu_{\rm B}$ obtained from a linear fit to the $1/\chi_{powder}(T)$ is in reasonable agreement with the $\mu_{\rm eff} = 10.35\ \mu_{\rm B}$ measured by Karunadasa *et al.*[22)]

The large differences in the Curie-Weiss temperatures observed for the three principal crystal directions of these compounds as well as the cross over from an easy-axis to an easy-plane type of anisotropy on cooling usually point to the presence of low-lying crystal field (CF) levels. Although there are preliminary reports on the position of the CF levels in the Sr$Ln_2$O$_4$ compounds[33,34)] based on inelastic neutron scattering measurements, there are no reports establishing the parameters of the CF Hamiltonians. The task of determining these parameters is complicated by the presence of the two crystallographically inequivalent sites for the rare-earth ions in the unit cell, both of them in a fairly low-symmetry environment.

### *3.1.2 Low temperature limit*

The results of low-temperature susceptibility measurements are summarised in Fig. 3 for fields of 100 Oe and 1 kOe applied along the $a$, $b$, and $c$ directions of the Sr$Ln_2$O$_4$ single crystals.

For SrEr$_2$O$_4$, the magnetic ordering transition at $T_N = 0.75$ K which has been reported previously from heat capacity and powder neutron diffraction measurements[23)] is seen most clearly for $H = 100$ Oe applied along the $c$ axis as a sharp drop in the value of $\chi(T)$ below 0.73 K (see top panel in Fig. 3). It is possible that a small variation in ordering temperature is due to a small difference in the stoichiometry of the samples: we noticed that at least some samples tend to lose weight if left unsealed for a long time. For the field applied along the $a$ axis, the susceptibility exhibits a broad maximum at temperatures slightly above $T_N$ with no sharp features, while the $\chi(T)$ curves are rather smooth and featureless for a field of 100 Oe applied along the $b$ axis (not shown).

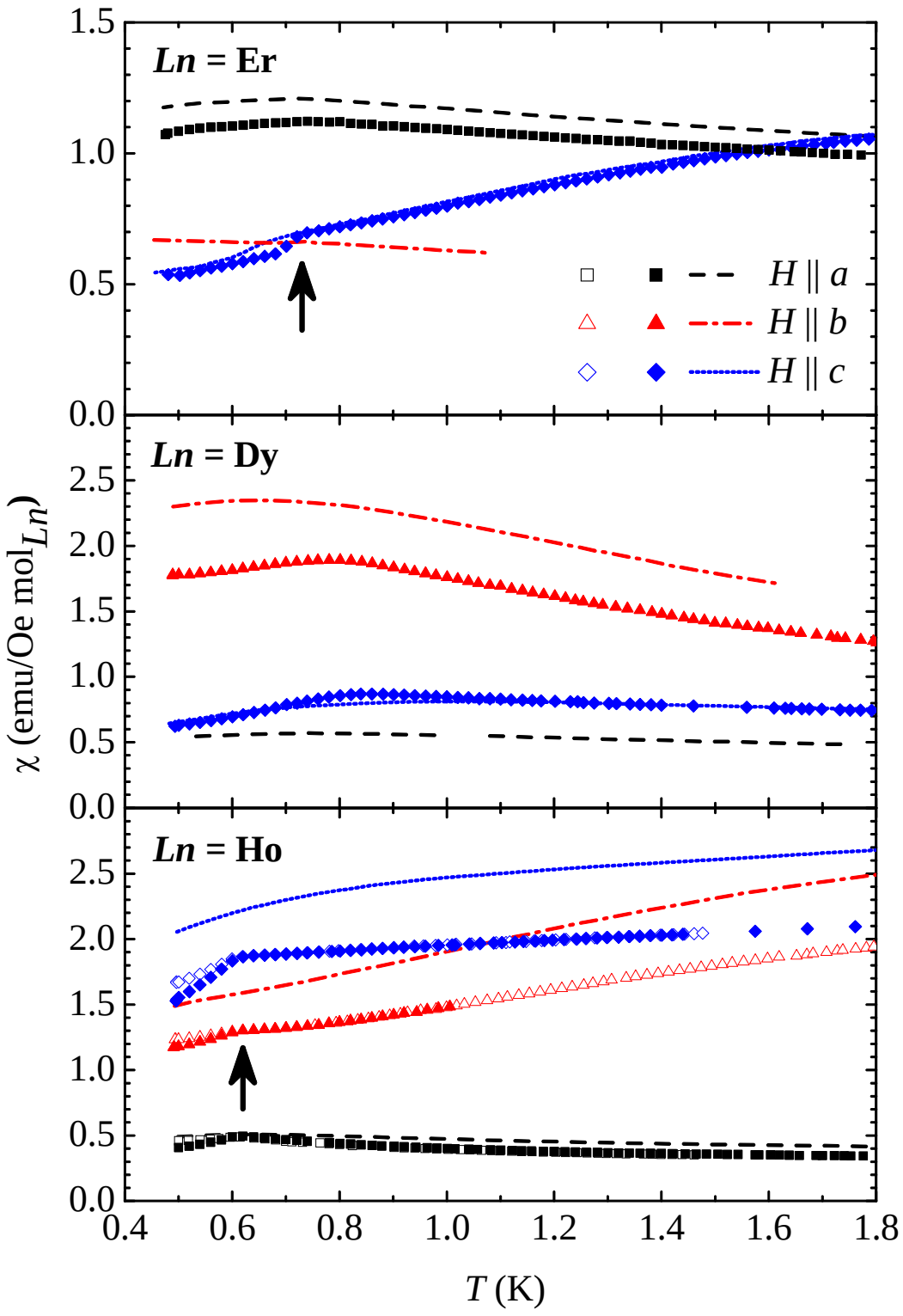

**Fig. 3.** (Colour online) Molar susceptibility curves for SrEr$_2$O$_4$, SrDy$_2$O$_4$ and SrHo$_2$O$_4$ obtained in the temperature range 0.48 to 1.8 K. The data for an applied field of 100 Oe are obtained on warming after field cooling (open symbols) or after zero-field cooling (solid symbols). Lines indicate the data for an applied field of 1 kOe obtained on warming after zero-field cooling. The arrows point to the anomalies associated with magnetic ordering transitions of 0.73 K in SrEr$_2$O$_4$ and 0.62 K in SrHo$_2$O$_4$.

For an applied field of 1 kOe the ordering transition in SrEr$_2$O$_4$ becomes almost invisible, as a sharp cusp in $\chi(T)_{H\|c}$ in 100 Oe is replaced with a much smoother curve, while the maximum in $\chi(T)_{H\|a}$ remains, but becomes slightly less pronounced. It is unclear whether the relatively modest applied field suppresses the magnetic ordering in SrEr$_2$O$_4$ or perhaps shifts it to much lower temperatures. Alternatively, the transition still exists in an applied field, but was not observed as the susceptibility simply became less sensitive to it.

A powder neutron diffraction study of SrEr$_2$O$_4$ points to an unusual magnetic behaviour, where only half of the Er sites (shown in either red or blue in Fig. 1) participate in the transition to Néel order[23)] by forming an antiferromagnetic structure parallel to the $c$ axis, while the other half of the magnetic sites remain disordered. Fur-thermore, recent single crystal neutron diffraction measurements[24] show evidence that this remaining half supports a separate shorter-range order (SRO) where the spins are aligned predominantly along the $a$ axis. This SRO structure coexists with the LRO structure at temperatures below $T_N$. One could imagine that both structures contribute differently towards the magnetic susceptibility, which might explain, in part at least, the highly anisotropic behaviour observed in $SrEr_2O_4$ at low temperature.

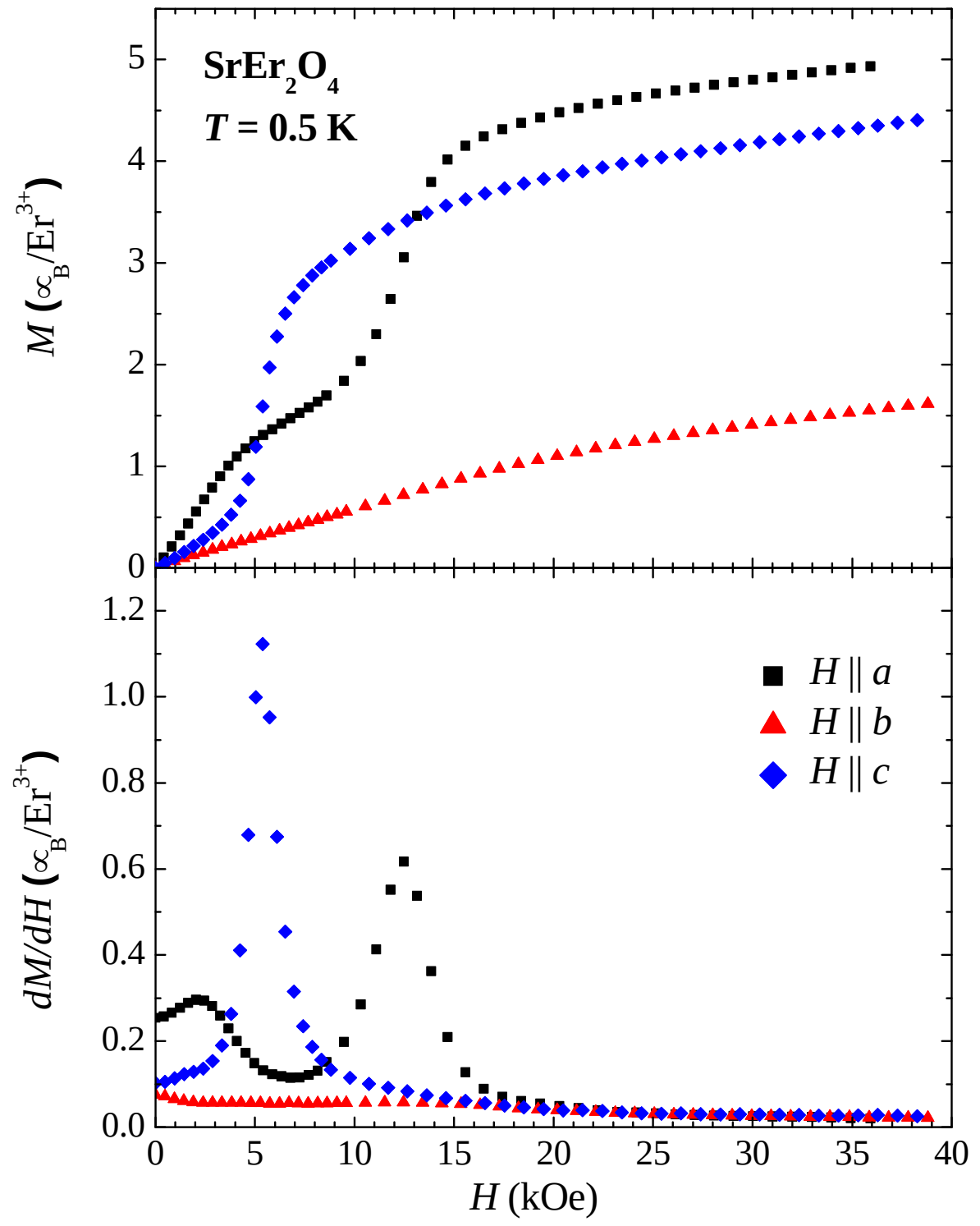

**Fig. 4.** (Colour online) (Top panel) Field-dependent magnetisation curves for $SrEr_2O_4$ obtained at 0.5 K in the range of fields 0 to 40 kOe. (Bottom panel) The field derivatives of the magnetisation.

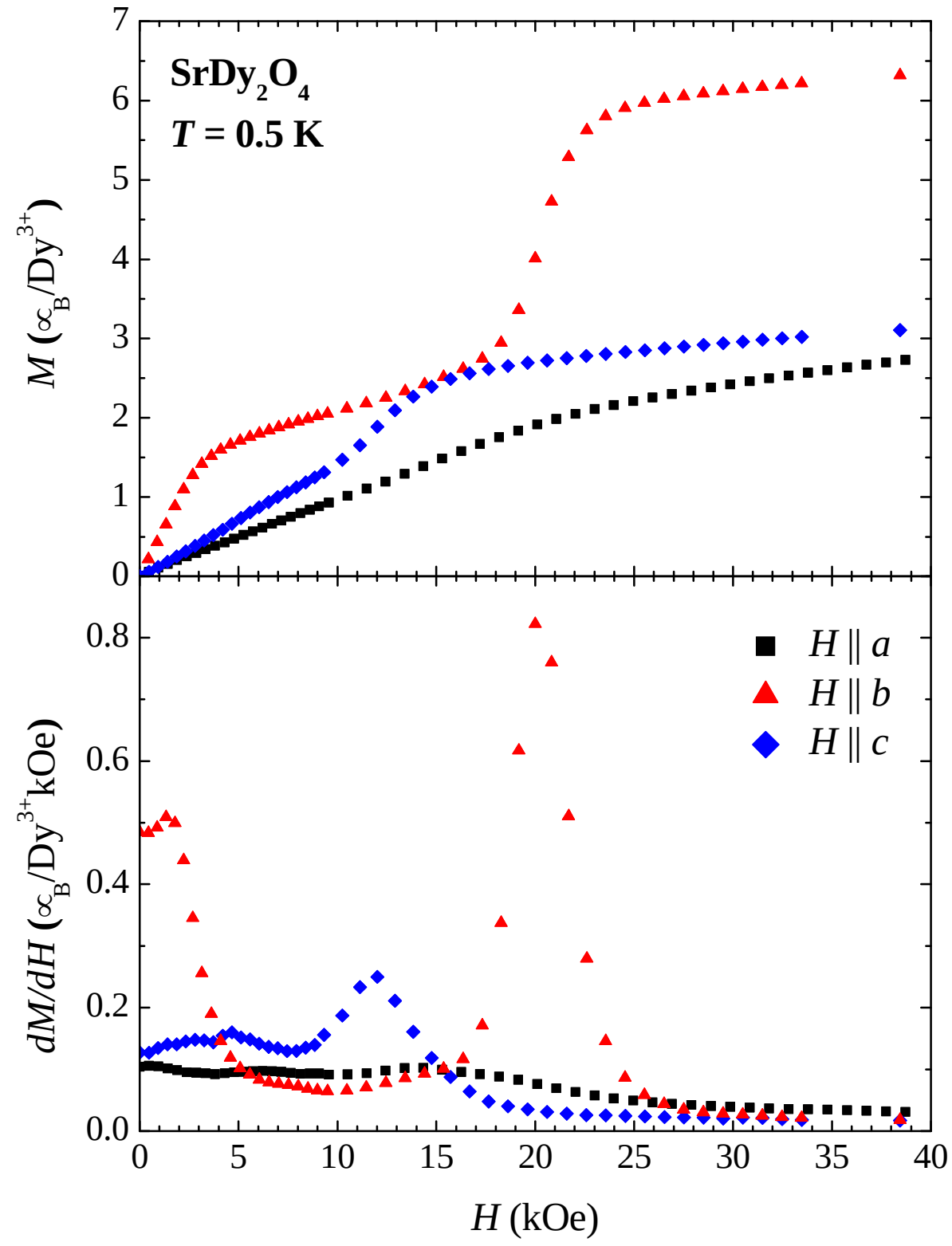

**Fig. 5.** (Colour online) (Top panel) Field-dependent magnetisation curves for $SrDy_2O_4$ obtained at 0.5 K in the range of fields 0 to 40 kOe. (Bottom panel) The field derivatives of the magnetisation.

For $SrHo_2O_4$, when a field of 100 Oe is applied along the different principal crystal axes, a cusp in the susceptibility at 0.62 K suggests the presence of a phase transition (Fig. 3, bottom panel). This is the first indication of a transition to a LRO state reported for this compound. Below the transition temperature, there is also a noticable difference between the data obtained on warming after cooling in field (FCW) and the zero-field-cooled warming (ZFCW) data. These data are included in the bottom panel of Fig. 3 as open (FCW) and closed (ZFCW) symbols. Remarkably, there is no noticeable difference between the FCW and the ZFCW data in $SrEr_2O_4$ at any temperature. In the absence of any data on the magnetic structure of $SrHo_2O_4$ below $T_N$ it is difficult to speculate whether the observed differences are due to incomplete order and glassiness or caused by magnetic domain wall motion.

Similar to what has been seen in $SrEr_2O_4$, a moderate field of 1 kOe applied to $SrHo_2O_4$ suppresses the cusp in the $\chi(T)$ curves (see bottom panel in Fig. 3). The higher-field data also appear to be almost insensitive to sample history.

The temperature dependent susceptibility curves for $SrDy_2O_4$ (see middle panel in Fig. 3) show some broad features at low temperatures indicative of SRO correlations, but the absence of sharp features down to 0.5 K for any field direction suggests that there is no transition to Néel order. In fact, our preliminary heat capacity and neutron scattering measurements on this compound[26] have revealed the lack of LRO down to 20 mK. Such a pronounced disparity between the ordering temperatures of isostructural rare-earth compounds is rare even among the highly-frustrated systems.

### *3.2 Field dependence of the magnetisation*

The field dependent magnetisation $M(H)$ and its derivative $dM/dH$, with $H$ applied along the $a$, $b$ and $c$ axes of the $SrLn_2O_4$ materials are shown in Figures 4, 5 and 6. Although the highest field measured was 70 kOe, we limit the field range in Figures 4-6 to 40 kOe, as above this field the magnetisation curves look rather featureless. No complete saturation of magnetisation is observed at any temperature in any of the compounds studied, as the $dM/dH$ values remain nonzero even at 70 kOe. This implies that the spins are still not fully aligned at this field – a result which is not surprising given the fact that the observed values of magnetisation remain much lower than expected from Hund's rules predictions for $Er^{3+}$, $Dy^{3+}$ and $Ho^{3+}$ ions and also that for one of the compounds, $SrEr_2O_4$, no saturation has been found at $T = 1.6$ K in a field as high as 280 kOe.[23)]

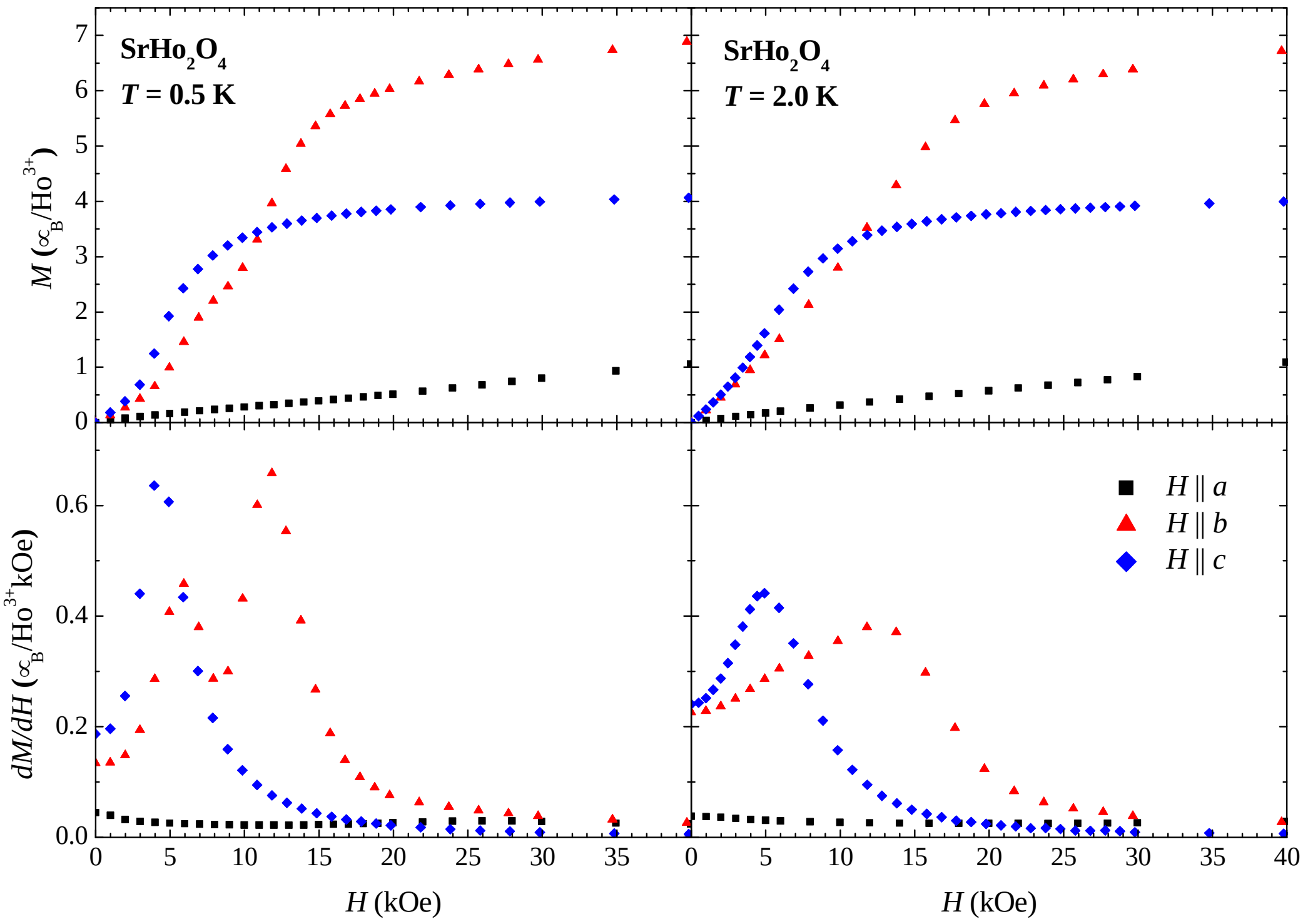

**Fig. 6.** (Colour online) (Top panels) Field-dependent magnetisation curves for $SrHo_2O_4$ obtained at 0.5 K (left) and 2.0 K (right) in the range of fields 0 to 40 kOe. (Bottom panels) The field derivatives of the magnetisation at 0.5 K (left) and 2.0 K (right).

In $SrEr_2O_4$, at 0.5 K (see Fig. 4) an easy-plane anisotropy can be seen, with multiple crossing points in the magnetisation measured for the field applied along the $a$ and $c$ axes. The $b$ axis is clearly the hard direction for magnetisation and the $dM/dH$ for a field applied along this direction remains small and nearly flat at all fields. The magnetisation process for a field applied within the easy $ac$ plane is, however, much more eventful.

For $H \parallel a$, an initial steep rise in magnetisation with a maximum in $dM/dH$ at $H_{c1} \approx 2$ kOe is followed by a much slower growth (with $dM/dH$ showing a minimum at $\approx 7$ kOe) and then by another sharp rise culminating at a maximum of $dM/dH$ at $H_{c2} \approx 12.5$ kOe. The average value of magnetisation on the plateau between $H_{c1}$ and $H_{c2}$ is 1.5 $\mu_B$, which equates to roughly a third of the value observed above $H_{c2}$. This is a clear sign of a field induced stabilisation of a collinear *uud* magnetic structure, in which two spins on each triangle are pointing up along the field with the third spin pointing down antiparallel to the field direction.[27,28] Compared to the $T = 1.5$ K data shown previously in Fig. 6 of ref.[23] the observed maxima in the $dM/dH$ curves at $T = 0.5$ K are sharper and the magnetisation slope between them is noticeably lower in value. This implies that on cooling below 0.5 K a further decrease in the $dM/dH$ value for $H_{c1} < H < H_{c2}$ is expected. It would be interesting to check whether the minimum in $dM/dH$ will effectively reach zero at $T \ll T_N$ or whether it will remain finite due to the presence of two independent magnetic structures in $SrEr_2O_4$.

For $H \parallel c$ the magnetisation process in $SrEr_2O_4$ is characterised by a single transition at $H_c \approx 5.4$ kOe with a very sharp maximum in the derivative $dM/dH$. The nature of this field-induced transition remains unknown at present.

In $SrDy_2O_4$, the field-dependent magnetisation curves at 0.5 K (see Fig. 5) also show multiple crossing points and lack of full saturation in high fields. Compared to $SrEr_2O_4$, a much more pronounced and extended plateau in magnetisation is seen between $H_{c1} \approx 1.6$ kOe and $H_{c2} \approx 20.3$ kOe for a field applied along the $b$ axis. The average magnetisation value on the plateau of 2.1 $\mu_B$ is again approximately a third of the value observed just above $H_{c2}$, justifying a claim for the stabilisation of a *uud* state. The derivative of magnetisation for a field applied along the hard $a$ axis remains almost flat at all fields. For $H \parallel c$ the magnetisation process in $SrDy_2O_4$ is dominated by a transition at $H_c \approx 12$ kOe with a maximum in the derivative $dM/dH$ that is not as sharp as in $SrEr_2O_4$. Another significantly less pronounced transition around 5 kOe cannot be ruled out, although the small feature seen in the $dM/dH$ at this field is close to the experimental resolution.

On warming the sample above 0.5 K (the data are not shown), the magnetisation plateau for $H \parallel b$ and the maximum in the $dM/dH$ for $H \parallel c$ gradually become less pronounced, while for the hard direction, $H \parallel a$, a small increase in $M(H)$ is found.

For $SrHo_2O_4$ (see Fig. 6) single crystal magnetisation measurements (top panel) and $dM/dH$ (bottom panel) shown for fields applied along the $a$, $b$, and $c$ directions at 0.5 K and 2 K also reveal the highly anisotropic nature of this rare-earth oxide. For $H \parallel a$ (clearly a hard magnetisation direction), $M(H)$ remains rather small in any field. For $H \parallel b$, a double phase transition is seen with $H_{c1} \approx 5.9$ kOe and $H_{c2} \approx 12$ kOe confining a narrow and not well defined plateau with the average value of magnetisation of 2.5 $\mu_B$. For $H \parallel c$ a single phase transition is seen at $H_c \approx 4$ kOe characterised by a sharp maximum in $dM/dH$.

On warming the sample to $T = 2.0$ K (see right-hand panels in Fig. 6), the plateau in the magnetisation disappears, the maximum in $dM/dH$ for $H \parallel c$ broadens and shifts to 5 kOe, while the $M(H)$ curve for $H \parallel a$ remains practically the same.

There are several key points currently impeding further comparative analysis of the magnetisation process in the $SrLn_2O_4$ compounds. Firstly, the low-temperature magnetic structures of $SrHo_2O_4$ and $SrDy_2O_4$ are as yet unknown. Current work on powder neutron diffraction[25] indicates that the LRO and SRO components may coexist in $SrHo_2O_4$, a scenario similar to that found in $SrEr_2O_4$.[24] For $SrDy_2O_4$ the situation could be radically different since no long-range order has been detected down to 20 mK.[26] Secondly, reliable data on the strength and the sign of the magnetic interactions in $SrLn_2O_4$ are required, as it is not clear whether the correct description of such systems should be based on a *2D honeycomb*, a *quasi-1D chain*, or a *zig-zag ladder* model. Lastly, a complete description of the CF levels would definitely be helpful in understanding the origins of the highly anisotropic behaviour found in these compounds as well as in identifying the roots of the dramatic differences exhibited in the magnetisation process.

## 4. Conclusions

To conclude, we present a comprehensive study of temperature dependence of the magnetic susceptibility and field dependence of the magnetisation performed on single crystal samples of geometrically frustrated magnetic materials $SrEr_2O_4$, $SrDy_2O_4$ and $SrHo_2O_4$. The measurements reveal the occurrence of a magnetic ordering transition in $SrHo_2O_4$ at 0.62 K and confirm that $SrEr_2O_4$ orders magnetically at about 0.73 K, while in $SrDy_2O_4$ such a transition is absent down to at least 0.5 K. Magnetisation plateaux corresponding to the field-induced stabilisation of the collinear *two-up one-down* structures are found in all three compounds. Apart from the plateaux, other field-induced transitions are also found at sufficiently low temperatures. We hope that this detailed information on the magnetisation process will contribute towards the development of theoretical models capable of describing the very complex low-temperature properties of the $SrLn_2O_4$ family and several similar honeycomb and triangles-based materials.

## Acknowledgments

The magnetometer used in this research was obtained through the Science City Advanced Materials project: Creating and Characterising Next Generation Advanced Materials, with support from Advantage West Midlands (AWM) and was part funded by the European Regional Development Fund (ERDF). The authors acknowledge financial support from the EPSRC, UK under grants EP/F02150X/1 and EP/E011802/1. We also thank M.R. Lees for a critical reading of the manuscript.